\documentstyle[graphicx,twoside,times]{jphysfr}
\def\dfrac#1#2{{\displaystyle\frac{#1}{#2}}}

\begin{document}
\setcounter{startpage}{1955}
\setcounter{lastpage}{1974}
\setcounter{vol}{4}
\setcounter{num}{11}
\setcounter{tome}{2}

\def\year{1994}
\def\month{11}

\shorttitle{ATOMIC DIFFRACTION BY A THIN PHASE GRATING}

\title{Atomic diffraction by a thin phase grating}

\author{C. Henkel, J.-Y. Courtois and A. Aspect}

\institute{%
Institut d'Optique Th\'eorique
  et Appliqu\'ee,\\ B.P. 147, 91403 Orsay Cedex, France}

\date{(Received 6 April 1994, accepted 1 June 1994)}

\maketitle

\begin{abstract}
We present a semiclassical perturbation method for the description of 
atomic diffraction by a weakly modulated potential. It proceeds in a way 
similar to the treatment of light diffraction by a thin phase grating, and 
consists in calculating the atomic wavefunction by means of action 
integrals along the classical trajectories of the atoms in the absence of 
the modulated part of the potential. The capabilities and the validity 
condition of the method are illustrated on the well-known case of atomic 
diffraction by a Gaussian standing wave. We prove that in this situation 
the perturbation method is equivalent to the Raman-Nath approximation, and 
we point out that the usually-considered Raman-Nath validity condition can 
lead to inaccuracies in the evaluation of the phases of the diffraction 
amplitudes. The method is also applied to the case of an evanescent wave 
reflection grating, and an analytical expression for the diffraction 
pattern at any incidence angle is obtained for the first time. Finally, 
the application of the method to other situations is briefly discussed.
%\\
%Classification:
%Physics Abstracts
%03.65S - 42.20G - 32.80P
%\\
\end{abstract}

\section{Introduction.}

The diffraction of atomic de Broglie waves by standing wave light fields 
[1-4]
or mechanical microstructures [5] has recently received considerable 
interest
because of its potential application as an atomic beam splitter, one of the
key components for the development of atom optics and interferometry
[6]. Among the different realizations of atom diffraction gratings 
proposed to
date, the near-resonant Kapitza-Dirac effect [7] (the diffraction of atom 
from
a standing laser field) has led by far to the most theoretical work
[8]. Widespread interest in this effect arose because the phenomenon is the
quantum mechanical analog of diffraction of light waves by a matter grating
and also because it is conceptually one of the simplest examples of 
stimulated
momentum transfer between atoms and light. The theoretical description of 
such
a transmission diffraction is greatly simplified in the Raman-Nath regime 
of
diffraction [1] where the change in kinetic energy of the atoms due to 
diffraction is neglected compared to the atom-field coupling. The problem 
then reduces to one dimension and the diffraction grating acts as a thin 
phase grating (the interaction between the atoms and the stationary laser 
wave only affects the phase of the atomic wavefunction). In contrast, the 
theoretical treatment of atomic diffraction by an evanescent wave 
reflection grating [3, 4] is 
known to lead to several unusual problems, which are related to the fact 
that the motion perpendicular to the diffraction grating can no longer be 
eliminated in a constant motion approximation. Indeed, the slowing down 
and eventually the reversal of the atomic motion is intrinsically 
associated with the reflection grating, and the problem remains 
necessarily two-dimensional. As a consequence, only a few attempts [3, 4] 
have been made to describe this diffraction process, and to our knowledge, 
no theoretical treatment provides a clear physical picture of atomic 
diffraction in the limit of small saturation of the atomic transition (a 
regime of great interest in atom optics experiments).

We present in this paper a semiclassical perturbation method which permits 
the treatment of atomic diffraction by a weakly modulated potential. The 
method applies both for transmission and reflection gratings, and proceeds 
in a way similar to the treatment of light diffraction by a thin phase 
grating. It is based on the evaluation of the atomic wavefunction by means 
of action integrals along the classical trajectories of the atoms 
calculated in the absence of the modulated part of the potential. In order 
to illustrate the capabilities and the limits of the perturbation method, 
we first consider the well-known near-resonant Kapitza-Dirac effect for 
which we prove the equivalence between our method and the Raman-Nath 
approximation. Furthermore, we show that the commonly-accepted validity 
condition of the latter actually leads to inaccuracies in the phases of 
the diffraction amplitudes. The case of atomic diffraction by a weakly 
modulated evanescent wave grating is then investigated and an analytical 
expression for the diffraction pattern at any incidence angle is derived 
for the first time. Finally, we discuss the application of the method to 
the situation of time-modulated potentials and to the case of multilevel 
atoms.

\section{Semiclassical perturbative calculation of the diffraction 
spectrum.}

We present in this section the principle of the semiclassical perturbation
method for calculating the diffraction spectrum of an atomic de Broglie 
wave
interacting with a weakly modulated potential. For illustration purposes, 
we
will here restrict the discussion to the case of a {\sl  spatially 
modulated\/} potential (the discussion of a time-modulated potential is 
postponed to section 5).

\subsection{Description of the model}
We consider the simple case of a two-level atom incident on the optical 
diffraction grating provided by an appropriate arrangement of laser beams. 
Because we are interested in the regime of coherent atom optics (limit of 
negligible spontaneous emission), we restrict ourselves to the case of low 
saturation of the atomic transition where the reactive part of the 
atom-laser wave coupling (light-shifts) is predominant over the 
dissipative part. We also assume that the detuning between the laser waves 
and the atomic frequency is properly chosen so that the atoms follow 
adiabatically the optical potential associated with the light-shifted 
ground-state level, and that any Doppler effect can be neglected. The 
Lagrangian of the atomic system is then of the form:

\begin{equation}
L({\bf r},{\bf \dot{r}})=L_0({\bf r},{\bf \dot{r}})-\varepsilon V({\bf r})
\end{equation}
\noindent
where $\bf r$ stands for the position of the atomic centre of mass, and 
$\bf
\dot{r}$ for its velocity. In equation (1), $\varepsilon V({\bf r})$ 
denotes
the spatially modulated part of the optical potential responsible for 
atomic
diffraction.This potential is assumed to be a small perturbation
($\varepsilon\ll 1$ is the perturbation parameter) compared to the non 
spatially modulated Lagrangian $L_0$ which contains the kinetic energy 
term.

\subsection{Calculation of the atomic wavefunction}

\subsubsection{The WKB method}
In the framework of a semiclassical (WKB) treatment of the atomic center of
mass motion, the atomic wavefunction is evaluated by means of action 
integrals
along classical 
\pagebreak

\noindent
atomic trajectories. 
As shown in Appendix A, assuming that 
at
time $t=t_{\rm i}$, the atoms have not yet interacted with the diffraction 
grating
and that the atomic wavefunction corresponds to a plane wave of momentum 
${\bf
  p}_{\rm i}$, the wavefunction at position ${\bf r}_{\rm f}$ and time $t_{\rm f}$ is given 
by:

\begin{equation}
\psi({\bf r}_{\rm f},t_{\rm f})=\exp\left[\frac{i}{\hbar}S({\bf r}_{\rm f},t_{\rm f}|{\bf 
p}_{\rm i},t_{\rm i})\right]
\end{equation}
\noindent
where:

\begin{equation}
S({\bf r}_{\rm f},t_{\rm f}|{\bf p}_{\rm i},t_{\rm i})={\bf p}_{\rm i}\cdot{\bf 
r}(t_{\rm i})+\int_{t_{\rm i}}^{t_{\rm f}}{\rm d}t
L[{\bf r}(t),{\bf\dot{r}}(t)]
\end{equation}
\noindent
is the action integral along the classical trajectory ${\bf r}(t)$ 
solution of the Euler-Lagrange equations associated with the Lagrangian 
$L$ (Eq.(1)), given the boundary conditions:

\begin{equation}
\left\{ \begin{array}{rcl}
 {\bf p}(t_{\rm i}) & = & {\bf p}_{\rm i} \\
{\bf r}(t_{\rm f}) & = & {\bf r}_{\rm f} \end{array} \right.
\end{equation} 
\noindent
with $M$ the atomic mass, and the requirement that ${\bf r}(t_{\rm i})$ be 
situated
outside the interaction region. Note that the first term in the right-hand
side of equation (3) takes into account the fact that the boundary 
conditions
(4) differ from the usual case where the initial and final {\sl  positions 
\/}
of the trajectory are specified. This term is associated with the phase of 
the
atomic wavefunction at the position ${\bf r}(t_{\rm i})$.

It is interesting to note that the above-described method for evaluating 
the
atomic wavefunction closely resembles the way one accounts for 
interference or
diffraction effects in conventional optics (where optical paths are 
calculated
along light rays derived from Fermat´s principle), and therefore is 
subject to
the same validity conditions. More precisely, it requires that both the
amplitude of the wavefunction and the optical potential vary slowly on the
scale of the atomic de Broglie wavelength. It thus breaks down near the 
points
where classical trajectories cross each other, e.g., near caustics or focal
points. Another characteristics of the WKB method is that it requires the
knowledge of the classical trajectories for the {\sl  total\/} Lagrangian 
$L$, which in practical application requires a numerical integration of 
the Euler-Lagrange equations.

\subsubsection{The perturbation method}
We show here that by taking advantage of the weakness of the spatially
modulated potential $\varepsilon V({\bf r})$, it is in fact possible to
evaluate the action integral (3) perturbatively up to leading order in
$\varepsilon$ using only the classical atomic trajectories for the 
unperturbed
Lagrangian $L$ [9, 10]. The interest of this method is that the unperturbed
trajectories are often known {\sl  analytically\/}, and therefore allow 
for analytical derivations of the diffraction spectrum. 

The perturbation method proceeds as follows. We expand the action integral 
(3)
and the actual atomic trajectories ${\bf r}(t)$, solutions of the 
Euler-Lagrange equations for the perturbed Langrangian $L$, in powers of 
the small parameter $\varepsilon$:

\begin{equation}
\left\{ \begin{array}{rcl}
 {\bf r}(t) & = &
{\bf r}_0(t)+\varepsilon{\bf r}_1(t)+\varepsilon^2{\bf
 r}_2(t)+\cdots \\ 
S({\bf r}_{\rm f},t_{\rm f}|{\bf p}_{\rm i},t_{\rm i}) & = &
S_0+\varepsilon
 S_1+\varepsilon^2 S_2+\cdots \end{array} \right.
\end{equation}
\noindent
Note that ${\bf r}_0(t)$ corresponds to the unperturbed classical 
trajectories, solutions of the equations of motion for the unperturbed 
Lagrangian $L_0$. Substituting equation (5) into (3), using equation (1) 
and
\pagebreak

\noindent
separating the different orders in $\varepsilon$, we get a set of 
equations the three first of which are (see Appendix A):

\begin{mathletters}
\begin{eqnarray}
S_0&=&{\bf p}_{\rm i}\cdot{\bf r}_0(t_{\rm i})+\int_{t_{\rm i}}^{t_{\rm f}}{\rm d}t L_0[{\bf
  r}_0(t),{\bf\dot{r}}_0(t)] 
\\
S_1&=&-\int_{t_{\rm i}}^{t_{\rm f}}{\rm d}t V[{\bf r}_0(t)] 
\\
S_2&=&-\frac{1}{2}\int_{t_{\rm i}}^{t_{\rm f}}\!{\rm d}t\,
{\bf r}_1(t)\cdot\nabla V[{\bf r}_0(t)]
\end{eqnarray}
\end{mathletters}

Equation (6a) is nothing but the exact action integral (3) in the limiting
case $\varepsilon=0$ where the modulated part of the optical potential
vanishes (no atomic diffraction). As a result, even though $S_0$ is 
generally
known analytically, it plays no role in the characterization of the
diffraction spectrum and therefore will not be considered in the
following. More relvant is equation (6b) which describes the phase-shift
accumulated by the atom along its {\sl  unperturbed\/} trajectory ${\bf
  r}_0(t)$, due to the presence of the modulated part of the optical
potential. $S_1$ actually contains the first order information about the
distortion of the atomic wavefront by the diffraction grating, and can 
thus be
used to derive the diffraction spectrum. This is the central point of our
perturbation treatment of atomic diffraction. It is clear however that 
such a
method will only be accurate in the limit of small $\varepsilon$, and we 
now
discuss its validity range. Consindering equations (2) and (5), it appears
that an appropriate condition is that $\varepsilon^2S_2$ be sufficiently 
small compared to $\hbar$, in other words that the higher perturbation 
orders do not significantly affect the atomic wavefunction. Using equation 
(6c), the validity condition of our perturbation method thus reads:

\begin{equation}
\frac{1}{2}\varepsilon^2\left|\int_{t_{\rm i}}^{t_{\rm f}}{\rm d}t{\bf r}_1(t)\cdot\nabla 
V[{\bf
    r}_0(t)]\right|\ll\hbar
\end{equation}
\noindent
for all possible trajectories. In fact, it is interesting to find an upper
limit for the left-hand side of equation (7) which permits a more 
transparent
physical interpretation of the validity condition. Note first that the
gradient $-\varepsilon\nabla V$ is the additional force acting on the atom 
due
to the modulated part of the potential. From the classical point of view, 
this
force is responsible for the momentum transfer involved in the diffraction
process. As a result, the time integral of the force will be of the order 
of
the maximum momentum transfer $\Delta p_{max}$ observed in the diffraction
spectrum (see Fig. 1). Second $|\varepsilon{\bf r}_1(t)|$ corresponds to 
the
deviation of the atom from its unperturbed trajectory. Its maximum is
$\Delta r_{max}$, the largest displacement observed after the atoms have 
excited the interaction region with the light grating (see Fig. 1). One 
thus obtains a condition for the validity of the perturbation method:

\begin{equation}
\frac{1}{2}\Delta p_{max}\Delta r_{max}\ll\hbar
\end{equation}
\noindent
which states that the error in the atomic phase due to the integration of 
the classical action along the unperturbed trajectory must be smaller than 
1, in other words that our approximate semiclassical estimate of the 
atomic wavefunction be essentially the same as in the WKB method. 

For illustration, let us consider the situation of a phase grating 
modulated along a single direction in space with a period $a$, and let 
$n_{max}$ be the maximum diffraction order observed in the experiment. 
Condition (8) then reads:

\begin{equation}
\Delta r_{max}\ll 2a/n_{max}
\end{equation}
\pagebreak

\noindent
which means that the deviation of the atoms from their unperturbed
trajectories must be small compared to the grating period {\sl  divided\/}
{\sl  by  the  maximum  diffraction  
order\/}. Note in particular that condition (9) is stronger than the 
validity condition of the WKB method, which requires that no caustics or 
focus points appear inside the interaction region between atoms and laser, 
and which reads:

\begin{equation}
\Delta r_{max}\ll a
\end{equation}
\noindent
As it happens, the validity condition (8) of the perturbation method 
embodies
the range of applicability of the semiclassical method, provided however 
that
the incident atomic de Broglie wavelength remains small compared to the
spatial variation scale of the optical potential. Finally, it is possible 
to
transport condition (8) from the spatial to the time domain by noticing 
that
$\Delta r_{max}$ is of the order of $\Delta p_{max}/M$ times the typical 
interaction time $\tau$ between the atoms and the light grating. In this 
way, one obtains a validity condition equivalent to (8) which reads:

\begin{equation}
\frac{\Delta p_{max}^2}{2M}\tau\ll\hbar
\end{equation}

\begin{figure}
\centerline{%
\resizebox{!}{84mm}{%
\includegraphics*{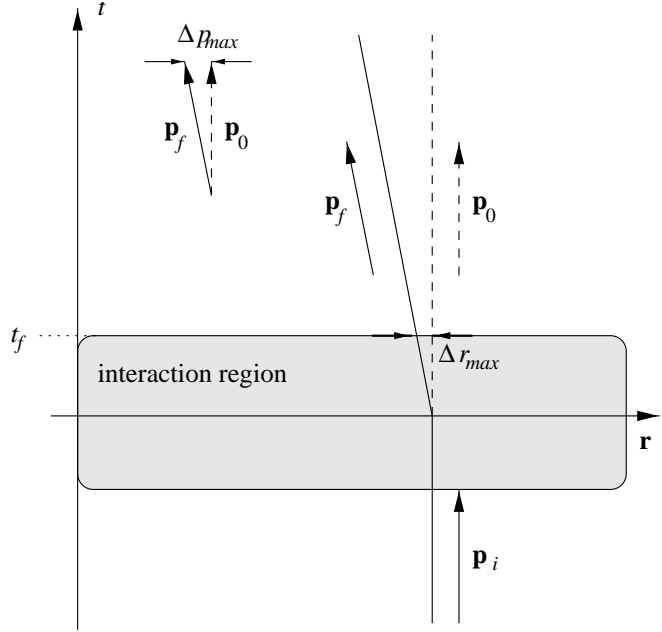}}}
\vspace*{-05mm}
\caption[]{%
Perturbed (solid line) {\sl  versus\/} unperturbed (dashed line)
atomic trajectories. At time $t=t_{\rm f}$ where the atoms exit the interaction
region with the diffraction grating, the maximum deviation in momentum
(resp. in position) between the perturbed and unperturbed atomic 
trajectories
is $\Delta p_{max}$ (resp. $\Delta r_{max}$).
}
\end{figure}

To conclude, we point out some important features of the above perturbative
method. First, the method clearly establishes an analogy between atomic
diffraction by a weakly modulated optical potential, and light diffraction 
by
a thin phase grating. Second, it is valid no matter the form of the
unperturbed atomic trajectories, and thus identically applies to 
transmission
and reflection gratings. Third, as will be shown in the following, it
generally leads to {\sl  analytical\/} expressions of the diffraction
amplitudes because $S_1$ only involves the {\sl  unperturbed\/} atomic 
trajectories which are often known analytically.

\subsection{Calculation of the atomic diffraction spectrum}
In a real experiment, atomic diffraction is observed at a distance from the
grating much larger than its typical spatial modulation scale (generally of
the order of a micron). As a result, the atomic wavefunction (2) evaluated
just at the exit of the interaction region with the grating does not 
properly
account for the observed diffraction pattern. In fact, it is necessary to
propagate the atomic wavefunction in the {\sl  far-field\/} region for 
obtaining the appropriate diffraction spectrum. Here are summarized the 
main results of Appendix B, where the detailed derivation of the 
diffraction spectrum is presented. 

For illustration purposes, we consider a two-dimensional geometry
($x$-$z$-plane) where atoms are incident from $z=-\infty$ with momentum ${\bf
  p}_{\rm i}$ on a diffraction grating with spatial modulation of period $a$ in 
the
$x$ direction (the formalism can be generalized straightforwardly to a
three-dimensional situation). Because of the periodicity of the grating, 
the
atomic diffraction spectrum displays a discrete pattern. The diffraction
orders are labeled by an integer number $n$ which denotes the momentum
transfer $n\hbar q$ (with $q=2\pi/a$) from the grating to the atoms in the
$x$ direction. It follows from energy conservation during the diffraction
process that the atomic momentum ${\bf p}^{(n)}=(p_x^{(n)},p_z^{(n)})$ 
associated with the $n$-th diffraction order reads:
\begin{equation}
\left\{ \begin{array}{rcl}
 p_x^{(n)} & = & 
p_{{\rm i},x}+n\hbar q 
\\ 
p_z^{(n)} & = & 
\sqrt{p_{\rm i}^2-(p_x^{(n)})^2} 
\end{array} \right.
\end{equation}
\noindent
where $p_{{\rm i},x}$ is the component of the incident atomic momentum along the 
$x$
axis. As shown in Appendix B, the diffraction amplitude $a_n$ associated 
with
the $n$-th diffraction order can be evaluated from the value of the atomic
wavefunction on a surface $z=z_{\rm f}$ located at the exit of the interaction 
region between the atoms and the light grating. More precisely, two 
situations can be distinguished depending on the way the atomic 
wavefunction is evaluated. 

\smallskip
\noindent
$\bullet$ {\sl Derivation from the WKB wavefunction.} 
--- 
In the case where the atomic wavefunction is evaluated
following the WKB method (Sect. 2.2.1), the diffraction amplitudes are 
given by (see Appendix B):
\begin{equation}
a_n=\frac{1}{2a}\int_0^a
{\rm d}x_{\rm f}\left[1+\frac{p_{{\rm f},z}(x_{\rm f})}{p_z^{(n)}}\right]\psi({\bf
  r}_{\rm f})\exp\left(-\frac{i}{\hbar}{\bf p}^{(n)}\cdot{\bf r}_{\rm f}\right)
\end{equation}
\noindent
where ${\bf r}_{\rm f} = (x_{\rm f},z_{\rm f})$ belongs to the ideally infinite 
surface $z=z_{\rm f}$ 
and
$p_{{\rm f},z}(x_{\rm f})$ denotes the $z$-component of the atomic momentum associated
with the classical perturbed trajectory ${\bf r}(t)$ satisfying the 
boundary
conditions (4). It is important to note that equation (13) does {\sl  not 
\/} merely correspond to the Fourier transform of the atomic wavefunction 
after the interaction with the diffraction grating. A difference indeed 
arises due to the supplementary factor in square brackets, which accounts 
for the angle of inclination of the trajectories with respect to the 
normal of the surface $z=z_{\rm f}$.

\smallskip
\noindent
$\bullet$ {\sl Derivation from the perturbation method.} 
--- 
In the case where the atomic wavefunction is 
evaluated following the perturbation method described in section 2.2.2, it 
is consistent to evaluate the diffraction amplitudes by substituting 
condition (11) into equation (13). As shown in Appendix B, this yields:
\begin{equation}
a_n=\frac{1}{a}\int_0^a {\rm d}x_{\rm f} \psi({\bf r}_{\rm f})\exp\left(-\frac{i}{\hbar}{\bf
    p}^{(n)}\cdot{\bf r}_{\rm f}\right)
\end{equation}
\noindent
which is the simple Fourier transform of the atomic wavefunction after the 
atom-grating interaction.

\pagebreak

\section{Illustration example: diffraction by a Gaussian standing wave.}

We illustrate in this section the capabilities of our perturbation method 
by considering the well-known case of atomic diffraction by a Gaussian 
standing wave, leading to the nearly-resonant Kapitza-Dirac effect. We 
show that our method recovers the analytical expression for the 
diffraction orders [1, 2] at any incident angle, and we prove its 
equivalence to the Raman-Nath approximation. The validity condition (8) of 
the method is verified by comparison with numerical WKB calculations, and 
we show that the accepted validity conditon of the Ranam-Nath 
approximation can lead to inaccuracies in the evaluation of the phases of 
the diffraction amplitudes.

\subsection{Calculation of the diffraction spectrum}

Consider a two-dimensional geometry where an atomic beam of momentum ${\bf
  p}_{\rm i}$ crosses the waist $w$ of a Gaussian standing wave at the incidence
angle $\theta$ ($\theta =0$ at normal incidence), and assume that the 
atomic kinetic energy is much larger than the height of the optical 
potential provided by the light field. The Lagrangian describing the atom 
dynamics thus takes the form (1) with:

\begin{mathletters}
\begin{eqnarray}
\displaystyle
L_0 & = & 
\frac{1}{2}M{\bf\dot{r}}^2 
\\
\varepsilon V({\bf r}) & = & 
\dfrac{\varepsilon}{\sqrt{2\pi}}V_1\exp(-2z^2/w^2)(1+\cos2kx) 
\end{eqnarray}
\end{mathletters}
\noindent
where $k=2\pi/\lambda$ is the wavevector associated with the laser 
wavelength
$\lambda$. As described in section 2.2.2, the perturbation method is based 
on
the evaluation of the action integral $S_1$ (Eq. (6b)) describing the
phase-shift undergone by the atoms along their unperturbed trajectories 
${\bf r}_0(t)=(x_0(t),z_0(t))$ which read:

\begin{equation}
\left\{ \begin{array}{rcl}
 x_0(t) & = & x_{\rm i}+\dfrac{p_{{\rm i},x}}{M}t 
\\
 z_0(t) & = & \dfrac{p_{{\rm i},z}}{M}t 
\end{array} \right.
\end{equation}
\noindent
The physical origin of the phase-shift can be visualized in figure 2a 
where we have projected the trajectories (16) onto the perturbation 
potenial (15b). More precisely, $S_1$ is evaluated by substituting 
equations (15b) and (16) into (6b). One thus gets:

\begin{equation}
S_1=-\frac{1}{2}V_1\tau(1+\beta_{KD}(\theta)\cos2kx_{\rm i})
\end{equation}
\noindent
where $\tau=Mw/p_{{\rm i},z}$ is the typical interaction time between the atoms 
and the laser wave, and $\beta_{KD}(\theta)$ is a real parameter 
describing the amplitude of the phase modulation which depends on the 
incidence angle $\theta$ [11]:

\begin{equation}
\beta_{KD}(\theta)=\exp\left(-\frac{1}{2}(kw\tan\theta)^2\right)
\end{equation}
\noindent
Note that the Kapitza-Dirac incidence factor $\beta_{KD}$ decreases
exponentially with the parameter $kw\tan\theta=k(p_{{\rm i},x}/M)\tau$, which
corresponds to the dimensionless displacement of the atoms along the 
standing
wave direction during the typical interaction time $\tau$. In particular, 
$\beta_{KD}=1$ (maximum value) at normal incidence, whereas $\beta_{KD}$ 
tends toward zero at grazing incidence (because of the spatial averaging 
of the potential modulation). 

\begin{figure}
\centerline{%
\resizebox{!}{175mm}{%
\includegraphics*{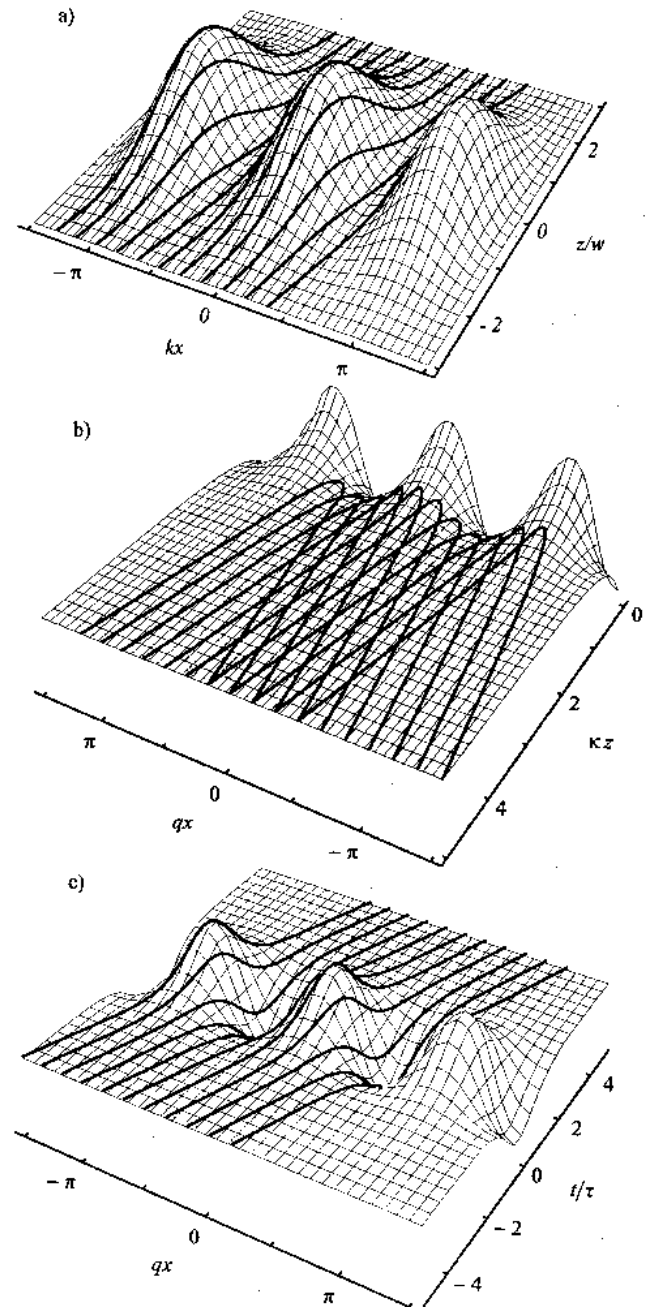}}}
\caption[]{%
Unperturbed atomic trajectories (bold lines) projected onto the
perturbation potential $\varepsilon V$(mesh). The atomic diffraction 
process results from the spatially modulated phase-shift undergone by the 
atoms along their unperturbed trajectories where they interact with the 
perturbation potential. (a) Case of a Gaussian standing wave. (b) Case of 
an evanescent wave reflection grating ($x-z$ coordinates). Note that the 
potential surface only corresponds to the spatially modulated part of the 
optical potential, and thus takes both positive and negatibe values. (c) 
Same as figure 2b, but $x$-time coordinates. Note the analogy with the 
case of figure 2a.
}
\end{figure}
\pagebreak

Finally, using equations (17), (5) and (2), one obtains:
\begin{equation}
\psi(x_{\rm f},z_{\rm f})\propto\exp\left[-\frac{i}{2\hbar}\varepsilon
  V_1\tau\beta_{KD}(\theta)\cos[2k(x_{\rm f}-z_{\rm f}\tan\theta)]\right]
\end{equation}
\noindent
from which the populations of the diffraction orders follow 
straightforwardly (see Eq. (14)):

\begin{equation}
|a_n|^2=J_n^2\left[\beta_{KD}(\theta)\frac{\varepsilon 
V_1\tau}{2\hbar}\right]
\end{equation}

\begin{figure}[hb]
\centerline{
\resizebox{!}{130mm}{%
\includegraphics*{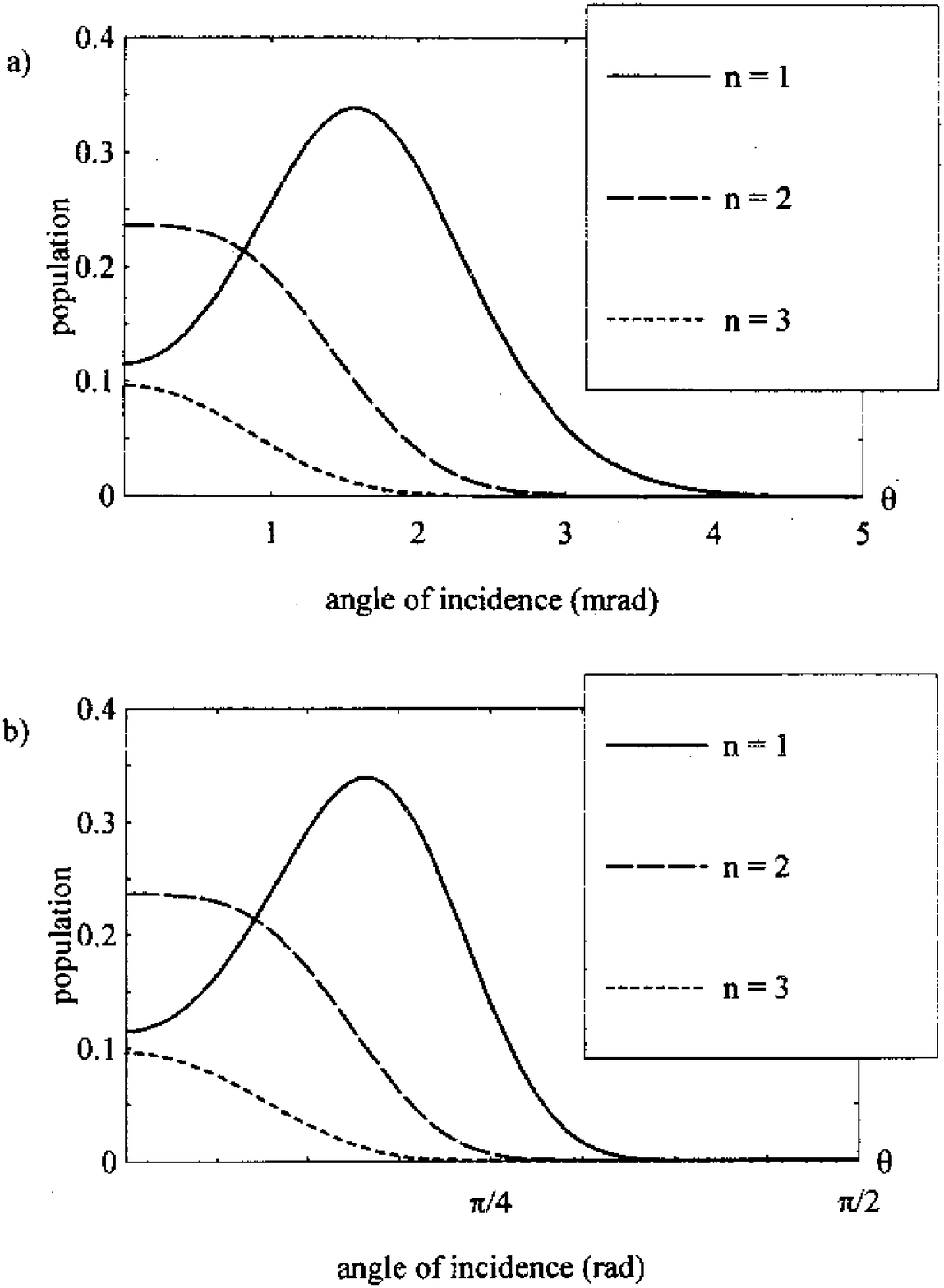}}}
\vspace*{-05mm}
\caption[]{%
Populations of the diffraction orders $n=1$-3 {\sl  versus\/} the
incidence angle $\theta$ (diffraction orders $n$ and $-n$ are equally
populated). (a) Case of a Gaussian standing wave transmission grating, with
$w=100\lambda$ and $\varepsilon V_1\tau/(2\hbar)=3$. The diffraction 
pattern
displays a dramatic dependence on $\theta$ on the scale of a few mrad. (b) 
By
contrast, the evanescent wave reflection grating leads to a diffraction
spectrum which varies much more smoothly with the incidence angle 
($\kappa=q$,
$\varepsilon p_{{\rm i},z}/(\hbar\kappa)=3$).
}
\end{figure}

\clearpage

\noindent
with $J_n$ the $n$-th Bessel function of integer order. Note that equation
(20) exactly corresponds to the result derived by Martin {\sl  et 
al.\/} [12] using the Raman-Nath approximation. Note also that the 
estimate
(20) is subject to the condition (11), which can be written in the form:

\begin{equation}
4n_{max}^2E_R\tau\ll\hbar
\end{equation}
\noindent
where $E_R=\hbar^2k^2/2M$ is the one-photon recoil energy, and $n_{max}$ is
the maximum diffraction order, given approximately by
$n_{max}\approx\beta_{KD}\frac{\varepsilon V_1\tau}{2\hbar}$.

We have represented in figure 3a the dependence of the populations of the
three first diffraction orders as a function of the angle of incidence
$\theta$, for a typical experimental value of the laser waist 
$w=100\lambda$
[12]. As reported in [12], the diffraction spectrum is found to display a 
dramatic sensitivity on the incidence angle on the scale of 5 mrad. This 
property results from the large value of $w$ used in figure 3a as well as 
in the experiment of reference [12] (see Eqs. (18) and (20).

\subsection{Comparison with the Raman-Nath approximation}

As shown above, the perturbation method allows us to recover exactly the
analytical expression for the populations of the diffraction orders 
obtained
using the Raman-Nath approximation. We show here that both approaches are
actually {\sl  equivalent\/} in the case of a Gaussian standing wave
diffraction grating. 

Let us briefly recall the main features of the Raman-Nath approximation [2,
12]. This approach is based on the Schr\"odinger equation describing the
interaction between fast atoms of incident momentum ${\bf p}_{\rm i}$ and the 
laser standing wave. In the interaction picture and after adiabatic 
elimination of the excited state, this equation reads:

\begin{equation}
i\hbar\partial_t\psi(x,t) = 
\left(-\frac{\hbar^2}{2M}\frac{\partial^2}{\partial
    x^2}+\varepsilon V[{\bf r}_0(t)]\right)\psi(x,t)
\end{equation}
\noindent
where $V$ is defined as in equation (15b) and ${\bf r}_0(t)$ as in equation
(16). The atomic wavefunction is then Fourier-expanded as 
$\psi(x,t)=\sum_n a_n(t)e^{2inkx}$, and the diffraction amplitudes $a_n$ 
are found to satisfy:

\begin{eqnarray}
i\dot{a}_n&=&\frac{\varepsilon}{2\sqrt{2\pi}}V_1\exp(-2t^2/\tau^2)\times\\
\nonumber 
&& {}\times (a_{n-1}e^{2ikp_{{\rm i},x}/Mt}+2a_n+a_{n+1}e^{-2ikp_{{\rm i},x}/Mt})
+4n^2E_Ra_n.
\end{eqnarray}
\noindent
Finally, the kinetic energy term in equation (23) is neglected compared to 
the
average atom-field coupling, provided that the Raman-Nath condition is
fulfilled:

\begin{equation}
4n_{max}^2E_R\ll\varepsilon\beta_{KD}V_1
\end{equation}
\noindent
Hence expression (20) is readily obtained.

In fact, the Raman-Nath approximation is equivalent to the Schr\"odinger
equation (22) {\sl  without  the  kinetic  energy
 term\/}, the solution of which:

\begin{equation}
\psi(x_{\rm f},t_{\rm f})\propto\exp\left(-\frac{i}{\hbar}\varepsilon\int_{t_{\rm i}}^{t_{\rm f}}{\rm d}t
  V[{\bf r}_0(t)]\right)
\end{equation}
\noindent
is rigorously equivalent to the result (19) obtained using the perturbation
method. It is important to note however that despite the mathematical
equivalence between the Raman-Nath approximation and the perturbation 
method,
both approaches are associated with different validity conditions. Indeed,
whereas condition (21) applies to the latter, the Raman-Nath approach is
associated with equation (24) which can be rewritten using the expression 
of
$n_{max}$ in the form:

\begin{equation}
2n_{max}E_R\tau\ll\hbar
\end{equation}
\noindent
Thus condition (21) is more severe than condition (26) by a factor of the
order of $n_{max}$, which can be significantly larger than 1. We show in 
the
following section that as far as the {\sl  populations\/} of the 
diffraction
orders are concerned, conditions (21) and (26) are not easily 
distinguishable,
but that equation (21) is actually more accurate than (26) for the 
estimate of
the {\sl  phases\/} of the diffraction amplitudes.

\subsection{Comparison with WKB calculations}
Here we discuss in more details the validity of the perturbation method, by
comparing it to the WKB treatment of atomic diffraction. In view of 
section 2,
one can express the WKB wavefunction $\psi_{{\rm WKB}}$ in terms of the perturbed
wavefunction $\psi_{{\rm pert}}$ as:

\begin{equation}
\psi_{{\rm WKB}}=\psi_{{\rm pert}}e^{i\delta\phi}
\end{equation}
\noindent
where in the limit of small deviations between the two approaches, 

\begin{equation}
\delta\phi(x_{\rm f})\approx -\eta\sin^2[2k(x_{\rm f}-z_{\rm f}\tan\theta)].
\end{equation}
\noindent
Here $\eta=4n_{max}^2E_R\tau/\hbar$ is the small parameter from condition
(21). As a result, from equation (27) and neglecting possible asymmetries 
in
the diffraction, the diffraction spectrum derived from the perturbation 
method
appears as the convolution of the WKB diffraction spectrum with the Fourier
transform of the function:
\begin{equation}
\exp(i\delta\phi)=e^{-i\eta/2}\sum_{m=-\infty}^{+\infty}
i^mJ_m(\eta/2)\exp[4imk(x_{\rm f}-z_{\rm f}\tan\theta)]
\end{equation}
\noindent
which only displays even diffraction orders, and has a typical width 
$\delta m
\approx \eta$.

In the limit $\eta\gg 1$ where condition (21) is not fulfilled, the
convolution spectrum is large and thus the diffraction spectrum derived 
using
the perturbation method differs significantly from the WKB spectrum. In
contrast, in the limit $\eta\ll 1$ where the validity condition (21) is 
fulfilled, the convolution spectrum is narrow, and hence the perturbation 
method is a good approximation. Quantitatively, the second order expansion 
of function (29) reads:
\begin{eqnarray}
\lefteqn{\exp(i\delta\phi)\approx} \\
 && \approx
 e^{i\eta/2}\left[1+i\frac{\eta}{2}\cos[4k(x_{\rm f}-z_{\rm f}\tan\theta)]-\frac{\eta^2}{8}\cos[8k(x_{\rm f}-z_{\rm f}\tan\theta)]-\frac{\eta^2}{16}\right]. 
\nonumber
\end{eqnarray}
\noindent
As far as the phases of the diffraction amplitudes are concerned, the
difference between the WKB and the perturbation method thus reduces to a
global phase-factor $e^{i\eta/2}$. It is also possible to compare the
populations of the diffraction orders using equation (30). After a
straightforward calculation, one finds:

\begin{equation}
|a_{n,{\rm pert}}|^2-|a_{n,{\rm WKB}}|^2\approx-\frac{\eta^2}{16}
\left[
2J_n(J_n+J_{n-4}+J_{n+4})-(J_{n-2}+J_{n+2})^2
\right]
\end{equation}
\noindent
where the argument of the Bessel functions is the same as in equation
(20). Since the term in square brackets is typically of the order of 
unity, we
see that the populations of the diffraction orders derived from the WKB and
the perturbation method only differ by a fraction of $\eta^2$. It is 
important
to note that the perturbation method, and hence the Raman-Nath 
approximation,
is therefore much more accurate for the evaluation of the {\sl  modulus\/}
{\sl  square\/} than for the estimate of the {\sl  phases\/} of the 
diffraction amplitudes. 

\pagebreak

We have performed numerical WKB calculations of the diffraction spectrum, 
and
compared the results with those of the perturbation method. We have 
confirmed
that in the case where condition (21) was fulfilled, the phases of the
diffraction amplitudes actually differ by the amount $-\eta/2$. The
populations of the diffraction orders were not clearly distinguishable as
expected from the smallness of $\eta^2$. We have been particularly 
interested
in the range of parameters where condition (26) was fulfilled, whereas
condition (21) was not. In the parameter space that we have explored, we 
have observed that the populations of the diffraction orders were still 
given to a good approximation by equation (20), but that the phases of the 
diffraction amplitudes actually differed significantly from the WKB 
estimates.

In conclusion, it turns out that the validity of both the perturbation 
method
and the Raman-Nath approximation are subject to one condition for the 
phases (Eq. (21)) and to another for the modulus square (Eq. (26)) of the 
diffraction amplitudes.

\section{Diffraction by an evanescent wave reflection grating.}

We consider in this section the atomic diffracton process associated with 
the
reflection grating provided by an evanescent wave having a small standing 
wave
component. Similar to the analogy between the Kapitza-Dirac effect and 
light
diffraction by an acoustic wave, one might expect the evanescent wave
reflection grating to be similar to the 
light diffraction grating produced by surface
acoustic waves (where the periodic undulations of the free surface act as a
surface grating). In fact, important differences arise between atom and
conventional optics concerning the reflection gratings. First, the 
repulsive
potentials of atomic mirrors typically vary on the scale of the optical
wavelength, which is generally much {\sl  larger\/} than the de Broglie
wavelength of the incident atoms. The situation is reversed in conventional
optics where metallic or dielectric surfaces achieve spatial changes in the
refractive index on a spatial scale much {\sl  smaller\/} than the optical
wavelength. Second, from a geometrical optics point of view, atomic
trajectories display properties different from those of light rays because 
of
the nonzero atomic mass. For instance, the fact that an atom can be
decelerated until its velocity is zero has no equivalent in conventional
optics. We show here that in the semiclassical regime of reflection, the
evanescent wave reflection grating is actually much more closely analogous 
to
the {\sl  transmission\/} than to the {\sl  reflection\/} grating of
conventional optics. More precisely, we prove that it is equivalent to the
transmission grating produced by a standing laser wave having an Eckart
profile ($\propto$ sech$^2$). Analytical expressions for the populations 
of the diffraction orders at any incidence angle are obtained, and the 
experimental conditions for observing the diffraction spectrum in the thin 
phase grating limit are briefly discussed.

\subsection{Calculation of the diffraction spectrum}
Consider a two-dimensional geometry where an ensemble of laser-cooled 
atoms of
momentum ${\bf p}_{\rm i}$ is incident on an evanescent wave reflection 
grating [3,
4] having a small standing wave component. Let $\theta$ denote the angle
between ${\bf p}_{\rm i}$ and the normal of the mirror, and assume that the 
optical
potential height provided by the evanescent wave is larger than the 
incident
kinetic energy of the atoms. The Lagrangian describing the atom dynamics 
thus
takes the form (1) with [13]:

\begin{mathletters}
\begin{eqnarray}
L_0 & = &
\frac{1}{2}M{\bf\dot{r}}^2-V_1e^{-2\kappa z}
\\
\varepsilon V({\bf r}) & = &
\varepsilon V_1e^{-2\kappa z}\cos2qx 
\end{eqnarray}
\end{mathletters}

\pagebreak
\noindent
where the wavevectors $\kappa$ and $q$ are typically of the order of the
vacuum wavevector $k=2\pi/\lambda$ associated with the laser wavelength
$\lambda$. Following the semiclassical perturbation method described in
section 2.2.2, we first derive the unperturbed atomic trajectories ${\bf
  r}_0(t)=(x_0(t),z_0(t))$ which are given by [14]:

\begin{equation}
\left\{ \begin{array}{rcl}
x_0(t) & = & x_{\rm i} + \dfrac{p_{{\rm i},x}}{M}t 
\\[3mm]
z_0(t) & = & 
-\dfrac{1}{2\kappa}\ln\left[
\dfrac{p_{{\rm i},z}^2}{2MV_1}\mbox{sech}^2(t/\tau)\right] 
\end{array} \right.
\end{equation} 
\noindent
where $\tau=M/\kappa p_{{\rm i},z}$ it the typical reflection time of the atom on
the evanescent wave mirror. Second, we evaluate the action integral $S_1$
(Eq. (6b)) by time-integration of the perturbation potential experienced by
the atoms along their unperturbed trajectories (see Fig. 2). Using 
equations
(32b) and (33), this potential is found to be:

\begin{equation}
\varepsilon 
V(t)=\varepsilon\frac{p_{{\rm i},z}^2}{2M}\cos[2q(x_{\rm i}+p_{{\rm i},x}/Mt)]
\mbox{sech}^2(t/\tau).
\end{equation}
\noindent
By comparison with equations (15b) and (16), it clearly appears that 
equation (34) is analogous to the perturbation potential experienced by 
the atoms in the Kapitza-Dirac geometry discussed in section 3 with the 
following substitutions:

\begin{equation}
\left\{ \begin{array}{l}
V_1\to\sqrt{2\pi}p_{{\rm i},z}^2/2M \\
k\to q \\
w\to 1/\kappa \\
\exp(-2z^2/w^2)\to \mbox{sech}^2(\kappa z).
\end{array} \right.
\end{equation}
\noindent
This shows that in the limit of a small standing wave component, the
evanescent wave {\sl  reflection\/} grating actually behaves as a {\sl 
transmission\/} grating (see Fig. 2). This is because the optical 
potential associated with the evanescent wave varies very smoothly on the 
scale of the incident de Broglie wavelenght [15].

Substituting equation (34) into expression (6b), one readily obtains:
\begin{equation}
S_1=-\beta_{EW}(\theta)\frac{p_{{\rm i},z}}{\kappa}\cos2qx_{\rm i}
\end{equation}
\noindent
where $\beta_{EW}(\theta)$ is the evanescent wave incidence parameter
analogous to equation (18) given by:
\begin{equation}
\beta_{EW}(\theta)=\frac{\pi\tan\theta \, q/\kappa}{
\sinh(\pi\tan\theta \, q/\kappa)}
\end{equation}
\noindent
which has a similar asymptotic behaviour as $\beta_{KD}$ (Eq.(18)) as 
$\theta$
tends toward 0 or $\pi/2$. Finally, using equations (36), (5), (2) and 
(14),
it is straightforward to derive the populations of the diffraction orders
which are found to read:
\begin{equation}
|a_n|^2=J_n^2\left(\varepsilon\beta_{EW}(\theta)
\frac{p_{{\rm i},z}}{\hbar\kappa}\right).
\end{equation}
\noindent
Note that equation (38) is analogous to expression (20) with the 
replacements
(35), as expected from the analogy with the Kapitza-Dirac effect. Note also
that the argument of the Bessel functions (or 
\pagebreak
\noindent
equivalently the number of
observable diffraction orders) is here proportional to the incident 
momentum in
the $z$ direction in the case of the reflection grating, whereas it is
inversely proportional to it in the case of the transmission grating (see
Eq. (20)). This is because in the case of the transmission grating, the 
height
of the perturbation potential is fixed and therefore the atomic phase-shift
decreases with incident momentum because of the $1/p_{{\rm i},z}$-dependence of 
the
interaction time $\tau$. By contrast, in the case of the reflection 
grating,
the height of the effective perturbation potential experienced by the 
atoms is
proportional to the {\sl  square\/} of the incident momentum, as shown by 
equation (35). For increasing momentum the increase of the potential 
height therefore overcomes the decrease of the interaction time, hence the 
phase-shift increases with $p_{{\rm i},z}$.

Similarly to the case of section 3, the accuracy of equation (38) is 
subject to condition (21) with 
$n_{max}\approx\varepsilon\beta_{EW}(\theta)p_{{\rm i},z}/\hbar\kappa$, with the 
restriction however that the semiclassical description of the reflection 
process be valid, i.e., $p_{{\rm i},z}\gg\hbar\kappa$ [14]. In fact, even though 
condition (21) is required for the accuracy of the phases of the 
diffraction amplitudes, equation (38) remains valid in a broader range of 
parameters given by a condition similar to equation (26).

We have represented in figure 3b the dependence of the populations of the 
three first diffracton orders as a function of the angle of incidence 
$\theta$, for typical experimental parameters. In contrast with the case 
of a Gaussian standing wave (Fig. 3(a)), the diffraction spectrum of the 
evanescent wave reflection grating does not exhibit a dramatic sensitivity 
on the incidence angle. This characteristic arises from the difference in 
the spatial extensions of the gratings along the $z$ direction 
($w=100\lambda$ for the transmisson grating, 
$w=1/\kappa\approx\lambda/2\pi$ for the reflection grating).

\subsection{Conditions for an experimental realization}
Here we briefly discuss the experimental conditions required for the
observation of atomic diffraction by an evanescent wave grating in the 
regime
of small modulation of the optical potential investigated in the preceding
section. Let us first examine the implications of the validity condition
(21). Using the expression of $\tau$ and assuming that $\kappa\approx k$, 
one
obtains as a first constraint:
\begin{equation}
p_{{\rm i},z}\gg 2n_{max}^2\hbar k.
\end{equation}
\noindent
In order to observe 5 diffraction orders, one thus has to achieve typically
$p_{{\rm i},z}\approx 100\hbar\kappa$. Second, one has to take care of the 
absence
of spontaneous emission events during the reflection process. As shown in
[16], the spontaneous emission probability $P_{sp}$ per reflection reads:
\begin{equation}
P_{sp}=\frac{\Gamma}{\Delta}\frac{p_{{\rm i},z}}{\hbar\kappa}
\end{equation}
\noindent
where $\Gamma$ is the natural width of the excited state and $\Delta$ is 
the
laser frequency detuning from resonance. In order to avoid spontaneous
emission, one thus needs a laser detuning $\Delta\approx 10^4\Gamma$. 
Third,
it is necessary to realize a sufficiently high optical potential barrier 
for
reflecting the atoms, i.e., $V_1\approx 10^4E_R$. Because $V_1$ is 
inversely
proportional to the detuning $\Delta$, this requires a large laser 
intensity,
typically of the order of $10^3\,$W/mm${}^2$ (case of rubidium atoms). This 
shows that such an experiment requires amplification techniques of the 
evanescent wave using either surface plasmons [17] or thin dielectric 
waveguides [18].

\section{Application of the method to other experimental situations.}
We have shown so far that the semiclassical perturbation method was a 
convenient tool for describing atomic diffraction by a weakly spatially 
modulated potential. We mention here possible extensions of the method to 
other situations of experimental interest.

\pagebreak
\noindent
$\bullet$ {\sl Atomic interferometry.} --- In atomic 
interferometry, it is often necessary to derive atomic phase-shifts due to 
small potentials, associated for example with a gravitational field or the 
rotation of the interferometer. In this situation, the expansion (6) can 
be used to calculate perturbatively the atomic phase-shifts using action 
integrals along the unperturbed atomic trajectories [19].
\\[2mm]
\noindent
$\bullet$ {\sl Time-modulated potential.} --- Because the 
central point of the perturbation method consists in evaluating the 
time-integral of a perturbation potential experienced by an atom along its 
unperturbed trajectory, the method applies straightforwardly to the 
situation where an atom interacts with an optical potential having a 
sufficiently small time-modulated component. For an illustration in the 
case of a time-modulated evanescent wave optical potential, see 
reference [20]. 
\\[2mm]
\noindent
$\bullet$ {\sl Multilevel atoms.} --- Certain experimental 
situations arise where the multilevel structure of the atoms plays an 
important role. For example, one can be interested in the atomic 
diffraction process by a Gaussian standing wave saturating the transition 
between a ground-state and a long-lived excited state. One can also 
consider a situation where the saturation of the optical transition is 
negligible, but where the Zeeman degeneracy of the ground-state is 
involved (e.g., an atom interacting with a standing wave displaying a 
polarization gradient). In such cases, the atomic state must be described 
by a spinor the components of which are associated with a given atomic 
internal state. However, in the case where the potential responsible for 
the mixing of the internal states can be considered as a small 
perturbation, it is possible to generalize our method and to derive the 
time-evolution of the atomic spinor by integration of the evolution 
operator associated with the perturbation potential along the unperturbed 
atomic trajectories [21].

\section{Conclusion.}

We have presented a semiclassical perturbation method which allows one to
describe in a simple way the interaction between an atom and a potential
having a small modulated component. This method is the analog in atom 
optics
of the treatment of light interaction with thin phase objects in 
conventinal
optics. It generally provides clear physical pictures as well as analytical
descriptions of the interaction process. It consists in evaluating the 
atomic
wavefunction by time-integration of the modulated potential along the
unperturbed classical trajectories of the atoms. A validity condition of 
the
method has been given and illustrated on the well-known Kapitza-Dirac 
effect,
where the method has been proved to be equivalent to the Raman-Nath
approximation. We have also used the perturbation method for deriving for 
the
first time an analytical expression for the populations of the diffraction
orders of an evanescent wave reflection grating at any incidence angle. 
This
perturbation method should prove interesting in a broader range of 
experimental situations, as shown by its application to the treatment of 
atom interaction with time-modulated potentials [20].

\section*{Acknowledgments.}

We are very grateful to C.J. Bord\'e, P. Grangier, R. Kaiser, N. 
Vansteenkiste, and C.I. Westbrook for many helpful and stimulating 
discussions.

%\appendix
\setcounter{equation}{0}
\renewcommand{\theequation}{A.\arabic{equation}}

\section*{Appendix A.}

\section*{Semiclassical perturbative derivation of the atomic wavefunction.}

In this appendix we derive equations (2) and (3) of section 2.1 for the 
semiclassical atomic wavefunction, as well as the perturbative expansion (6) 
of the integral (3).

\subsection*{A.1 Semiclassical derivation of the atomic wavefunction}
For a time-independent Hamiltonian system, the quantum propagator can be
represented by a Feynman path integral [22]. Thus, if at time $t=t_{\rm f}$, the
atom is described by a plane wave of momentum ${\bf p}_{\rm i}$:
\begin{equation} 
\psi({\bf r},t_{\rm i}) = \exp\frac{i}{\hbar}({\bf
      p}_{\rm i}\cdot{\bf r}) 
\end{equation}
then at time $t=t_{\rm f}$, the atomic wavefunction is given by the convolution 
product:
\begin{equation} 
\psi({\bf r}_{\rm f},t_{\rm f})=\int d{\bf r}_{\rm i} K({\bf
      r}_{\rm f},t_{\rm f}|{\bf r}_{\rm i},t_{\rm i})\exp\left(\frac{i}{\hbar}{\bf p}_{\rm i}\cdot{\bf
        r}_{\rm i}\right) 
\end{equation}
with the Feynman propagator defined as the path-integral:

\begin{equation} K({\bf r}_{\rm f},t_{\rm f}|{\bf 
r}_{\rm i},t_{\rm i})=\int_{({\bf r}_{\rm i},t_{\rm i})}^{({\bf r}_{\rm f},t_{\rm f})}{\cal D}[{\bf 
r}(t)]\exp\frac{i}{\hbar}\left(\int_{t_{\rm i}}^{t_{\rm f}} L({\bf r}(t),\dot{\bf 
r}(t)){\rm d}t\right). \end{equation}
\noindent                                                                      
The measure ${\cal D}[{\bf r}(t)]$ signifies that the integration is to be
taken over all trajectories ${\bf r}(t)$ satisfying the boundary 
conditions:
\begin{equation}
\left\{ \begin{array}{rcl}
{\bf r}(t_{\rm i})& = &{\bf r}_{\rm i} \\
{\bf r}(t_{\rm f})& = &{\bf r}_{\rm f}
\end{array} \right.
\end{equation}

The semiclassical version of the quantum propagator (A.2) arises from a
stationary-phase approximation. The path integral is then dominated by
contributions from classical trajectories since these render the phase of 
the
integrand stationary. The phase of the atomic wavefunction (times $\hbar$) 
is
therefore given by the value of the generalized action:

\begin{equation}
S[{\bf r}(t),{\bf r}_{\rm i}]={\bf p}_{\rm i}\cdot{\bf r}_{\rm i}+\int_{t_{\rm i}}^{t_{\rm f}} L({\bf
  r}(t),\dot{\bf r}(t)) {\rm d}t \end{equation}
\noindent
for the specific initial point ${\bf r}_{\rm i}$ and the trajectory ${\bf r}(t)$
which fullfill the stationary-phase condition:

\begin{equation}
0=\delta S={\bf p}_{\rm i}\cdot\delta{\bf r}_{\rm i}+\delta{\bf 
r}\cdot\left.\frac{\delta
  L}{\delta \dot{\bf r}}\right|_{t_{\rm i}}^{t_{\rm f}}+\int_{t_{\rm i}}^{t_{\rm f}} \delta{\bf
r}\cdot\left(\frac{\partial L}{\partial{\bf r}}-\frac{\rm d}{{\rm d}t}\frac{\partial
  L}{\partial\dot{\bf r}}\right){\rm d}t \end{equation}
\noindent
for any small deviation ($\delta{\bf r}_{\rm i}$,$\delta{\bf r}(t)$) from the 
path of stationary phase. Because the boundary conditions (A.4) impose the 
relation:

\begin{equation}
\delta{\bf r}(t_{\rm i})=\delta{\bf r}_{\rm i} \end{equation}
\noindent
it follows from (A.6) that:

\begin{equation}
\left\{ \begin{array}{l}
{\bf p}(t_{\rm i})=\dfrac{\partial L}{\partial\dot{\bf r}}(t_{\rm i})={\bf p}_{\rm i} 
\\[3mm]
\dfrac{\partial L}{\partial {\bf r}}-\dfrac{\rm d}{{\rm d}t}\dfrac{\partial 
L}{\partial\dot{\bf r}}=0.
\end{array} \right.
\end{equation}
\noindent
As expected, one finds that the stationary-phase trajectory ${\bf r}(t)$ 
is a
{\sl  classical\/} trajectory (it is solution of the Euler-Lagrange 
equations of motion) which satisfies the boundary conditions (4). The 
semiclassical atomic wavefunction is hence given by equations (2) and (3) 
of section 2.1.

\subsection*{A.2 Perturbative expansion of the action integral}
We consider the method described in section 2.2 for deriving the action
integral (3) by an expansion in powers of the small parameter
$\varepsilon$. In fact, only the second-order term is non-trivial: the 
zero-th
order term $S_0$ corresponds to the absence of the modulated potential and
hence to expression (6a), and equation (6b) arises from the fact that the 
term
linear in $\varepsilon$ involving the nonperturbed Lagrangian $L_0$ 
vanishes
by virtue of the stationary phase-condition (A.6) for the unperturbed
trajectory ${\bf r}_0(t)$. In order to calculate the second-order term 
(6c),
we have to expand both the boundary conditions (4) and the Euler-Lagrange
equations of motion for the {\sl  perturbed\/} trajectory up to first 
order
in $\varepsilon$. This yields the relations:

\begin{equation}
\left\{ \begin{array}{l} 
0=\left(\dot{\bf r}_1\cdot\dfrac{\partial}{\partial\dot{\bf r}}+{\bf
    r}_1\cdot\dfrac{\partial}{\partial{\bf r}}\right)\left.\dfrac{\partial
    L_0}{\partial\dot{\bf r}}\right|_{{\bf r}={\bf r}_0, \, \dot{\bf 
r}=\dot{\bf
    r}_0, \, t=t_0} \\ \\
0=\left(\dot{\bf r}_1\cdot\dfrac{\partial}{\partial\dot{\bf r}}+{\bf
    r}_1\cdot\dfrac{\partial}{\partial{\bf r}}\right)\dfrac{\partial
  L_0}{\partial{\bf r}}-\dfrac{\partial V}{\partial{\bf 
r}}-\dfrac{\rm d}{{\rm d}t}\left(\dot{\bf r}_1\cdot\dfrac{\partial}{\partial\dot{\bf 
r}}+{\bf
    r}_1\cdot\dfrac{\partial}{\partial{\bf r}}\right)\left.\dfrac{\partial
    L_0}{\partial\dot{\bf r}}\right|_{{\bf r}={\bf r}_0,\, \dot{\bf 
r}=\dot{\bf
    r}_0} \end{array}
 \right.
\end{equation}
\noindent
The second-order term of the action is equal to:
\begin{equation}
S_2 = 
{\bf p}_{\rm i}\cdot{\bf r}_2(t_{\rm i})+\int_{t_{\rm i}}^{t_{\rm f}} {\rm d}t 
%\times 
%\nonumber
%\\ &&
%{}\times 
\left(
{\bf r}_2\cdot\dfrac{\partial L_0}{\partial{\bf r}}+\dot{\bf
    r}_2\cdot\dfrac{\partial L_0}{\partial\dot{\bf r}}-{\bf
    r}_1\cdot\dfrac{\partial V}{\partial{\bf r}}+\dfrac{1}{2}\left({\bf
      r}_1\cdot\dfrac{\partial}{\partial{\bf r}}+\dot{\bf
      r}_1\cdot\dfrac{\partial}{\partial\dot{\bf r}}\right)^2 L_0
\right).
%\end{eqnarray}
\end{equation}
\noindent
Note that all the quantities inside the integral are evaluated along the 
{\sl 
unperturbed\/} trajectory. Finally, using (A.9), the Euler-Lagrange 
equation
for ${\bf r}_0(t)$ and integrations by parts, equation (A.10) is readily 
simplified to get the result (6c).

\setcounter{equation}{0}
\renewcommand{\theequation}{B.\arabic{equation}}

\section*{Appendix B.}
\section*{Calculation of the diffraction spectrum.}

In this appendix we derive the expressions of the atomic diffraction 
amplitudes (13) and (14).

\subsection*{B.1 Quantum-mechanical integral theorem of Helmholtz and 
Kirchhoff}
In a potential-free region of space, the wavefunction of an atom of kinetic
energy $E$ satisfies the Schr\"odinger equation:
\begin{equation}
\nabla^2\psi+\frac{2ME}{\hbar^2}\psi=0.
\end{equation}
\noindent
Because equation (B.1) has the same form as the Helmholtz equation in
electromagnetic theory, it is possible to use a quantum-mechanical version of
the integral theorem of Helmholtz and Kirchhoff [23] to express the atomic
wavefunction in the far-field region (which describes the diffraction 
pattern)
from its value on a boundary surface $\Sigma$ located in the free-field 
region
reached by the atoms after their interaction with the diffractron grating. 
One
thus has:

\begin{equation}
\psi({\bf r})_{\rm farfield}=\frac{1}{4\pi}\int d\Sigma 
\, {\bf n}\cdot[G_E({\bf
  r},{\bf r}_{\rm f})\nabla\psi({\bf r}_{\rm f})-\psi({\bf r}_{\rm f})\nabla G_E({\bf r},{\bf
  r}_{\rm f})] \end{equation}
\noindent
where ${\bf n}$ denotes the outward-pointing normal to the surface 
$\Sigma$,
and ${\bf r}_{\rm f}$ corresponds to the endpoints of the classical trajectories
along which the action integral (3) is calculated. The integral in equation
(B.2) involves on one hand the quantum propagator $G_E({\bf r},{\bf r}_{\rm f})$ 
for
a free particle of kinetic energy $E$, solution of:

\begin{equation}
\left[\nabla^2+\frac{2M}{\hbar^2}E\right]G_E({\bf r},{\bf
  r}_{\rm f})=-4\pi\delta({\bf r}-{\bf r}_{\rm f}) 
\end{equation}
\noindent
($\delta$ is the Dirac delta function), and on the other hand the value of 
the
atomic wavefunction and its spatial derivative ${\bf n}\cdot\nabla\psi$ on 
the
surface $\Sigma$. More precisely, $\nabla\psi$ is evaluated either from the
WKB or the perturbation method using the relation:

\begin{equation}
\nabla\psi({\bf r}_{\rm f})=\frac{i}{\hbar}{\bf p}_{\rm f}({\bf r}_{\rm f})\psi({\bf r}_{\rm f})
\end{equation}
\noindent
where ${\bf p}_{\rm f}({\bf r}_{\rm f})$ is the local atomic momentum at the endpoint
point ${\bf r}_{\rm f}$ of the classical trajectory. Note that whereas the 
direction
of ${\bf p}_{\rm f}({\bf r}_{\rm f})$ generally depends on ${\bf r}_{\rm f}$, its modulus is
constant, equal to that of the incident atomic momentum ${\bf p}_{\rm i}$ 
(because of energy conservation in the diffraction process).

\subsection*{B.2 Diffraction by a two-dimensional periodic grating}
We consider here the particular case of a two-dimensional diffraction 
grating located in the $xOz$ plane (or equivalently a three-dimensional 
grating with translational invariance along the $y$-axis). We assume that 
the grating has a finite extent in the $z$-direction, but is infinite in 
the $x$-direction along which it is spatially modulated with the 
periodicity $a$. Using the notations of section 2, one thus has:

\begin{equation}
\forall x, \; \; V(x+a)=V(x) \end{equation}
\noindent
It follows from equation (B.5) and the Bloch theorem that the atomic 
wavefunction will take the form:

\begin{equation}
\forall x,z \; \; \psi(x+a,z)=e^{ip_{{\rm i},x}a/\hbar}\psi(x,z) 
\end{equation}
\noindent
with $p_{{\rm i},x}$ the component of the incident atomic momentum along the
$x$-axis [24]. In order to take advantage if this property in the 
derivation
of the diffraction spectrum, it is convenient to define the surface 
$\Sigma$
of equation (B.2) as a line $z=z_{\rm f}$ (${\bf n}=-{\bf e}_z$ is thus 
independent
of ${\bf r}_{\rm f}$) and to express the two-dimensional quantum propagator 
$G_E({\bf
  r},{\bf r}_{\rm f})$ in the free-field region ($z>z_{\rm f}$) in the form [23]:

\begin{equation}
G_E({\bf r},{\bf r}_{\rm f})=i \int {\rm d}p_x 
\frac{1}{p_z}\exp\left(\frac{i}{\hbar}{\bf
    p}\cdot({\bf r}-{\bf r}_{\rm f})\right) \end{equation}
\noindent
where $p_z=\sqrt{2ME-p_x^2}$ (the integration is restricted to the interval
$|p_x|\le\sqrt{2ME}$ because the evanescent wave components of the 
proagator
do not contribute to the far-field wavefunction). Using equations (B.4), 
(B.6)
and (B.7), and dividing the integration range of equation (B.2) into 
intervals
of length $a$, one obtains:
\begin{eqnarray}
\psi({\bf r})_{\rm farfield}
&=&
\frac{1}{4\pi\hbar}\sum_{n=-\infty}^{+\infty}\frac{1}{a}\int_0^a 
{\rm d}x_{\rm f} \int {\rm d}p_x \left[1+\frac{p_{{\rm f},z}(x_{\rm f})}{p_z}\right]\psi(x_{\rm f},z_{\rm f}) 
\times{}
\nonumber
\\
& & {}\times 
\exp\left(\frac{i}{\hbar}n(p_{{\rm i},x}-p_x)a\right)
\exp\left(\frac{i}{\hbar}{\bf
   p}\cdot({\bf r}-{\bf r}_{\rm f})\right). \end{eqnarray}

\pagebreak
\noindent
Using the relation:
\begin{equation}
\sum_{n=-\infty}^{+\infty}\exp\frac{i}{\hbar}n(p_{{\rm i},x}-p_x)a=\hbar
q\sum_{n=-\infty}^{+\infty} \delta(p_x-p_{{\rm i},x}-n\hbar q)
\end{equation}
\noindent
and expression (12) for the momenta associated with the different 
diffraction orders, equation (B.8) finally yields:

\begin{equation}
\psi({\bf r})_{\rm farfield}=\sum_{n=-\infty}^{+\infty}
a_n\exp\left(\frac{i}{\hbar}({\bf p}^{(n)}\cdot{\bf r})\right)
\end{equation}
\noindent   
where $a_n$ is the diffraction amplitude associated with the n-th 
diffraction order, which reads:

\begin{equation}
a_n=\frac{1}{2a}\int_0^a
{\rm d}x_{\rm f}\left[1+\frac{p_{{\rm f},z}(x_{\rm f})}{p_z^{(n)}}\right]\psi({\bf
  r}_{\rm f})\exp\left(-\frac{i}{\hbar}{\bf p}^{(n)}\cdot{\bf r}_{\rm f}\right).
\end{equation}

Even though equation (B.11) may be used to derive the atomic diffraction
spectrum in the framework of the perturbation method, it seems more 
consistent
to retain only the terms of (B.11) which correspond to the accuracy range 
of
the method. As previously discussed in section 2.2.2 (Eq.(11)), in the
validity domain of our semiclassical method one has:

\begin{equation}
\frac{\Delta p_{max}^2}{p_{\rm i}^2}\ll\frac{\hbar}{E\tau}\ll1
\end{equation}
\noindent
and hence:
\begin{equation}
1-\left|\frac{p_z}{p_{\rm i}}\right|,
1-\left|\frac{p_z^{(n)}}{p_{\rm i}}\right|\ll1-\sqrt{1-\frac{\Delta
      p_{max}^2}{p_{\rm i}^2}}\approx\frac{\Delta p_{max}^2}{2p_{\rm i}^2}\ll1
\end{equation}
\noindent
which yields:
\begin{equation}
\left|\frac{p_{{\rm f},z}}{p_z^{(n)}}-1\right|\ll1
\end{equation}
\noindent
Finally, one obtains the expression of the diffraction amplitude for the 
atomic wavefunction derived using the perturbation method:

\begin{equation}
a_n=\frac{1}{a}\int_0^a {\rm d}x_{\rm f} \psi({\bf r}_{\rm f})\exp\left(-\frac{i}{\hbar}{\bf
    p}^{(n)}\cdot{\bf r}_{\rm f}\right)
\end{equation}
\noindent
which is merely the Fourier transform of the wavefunction evaluated on the 
line $z=z_{\rm f}$, after interaction with the diffraction grating.


\begin{thebibliography}{99}
\bibitem{[13]} We have neglected the distortion of the optical potential
due to van der Waals interaction between the atom and the mirror surface.
\bibitem{[15]} Note that this property would not hold in the quantum 
regime of reflection achieved for incident atomic momentum of the order of 
$\hbar\kappa$, and for which the evanescent wave grating would behave as a 
hard potential barrier similar to a reflection grating in conventional 
optics [14].
\end{thebibliography}
\end{document}